\documentclass[twocolumn,showpacs,preprintnumbers,amsmath,amssymb,superscriptaddress,graphicx,graphics,pstricks]{revtex4}
\usepackage{amsmath}
\usepackage{graphicx}
\usepackage[latin1]{inputenc}
\usepackage{pstricks}
\usepackage{pst-node}
\usepackage{pst-coil}
\usepackage{pst-grad}
\usepackage{multido}
\usepackage{dcolumn}
\usepackage{bm}

\begin{document}

\title{Thinking anew causality problems for the radiation reaction force}

\author{Reinaldo de Melo e Souza}
\affiliation{Instituto de Fisica, UFRJ, CP 68528, Rio de Janeiro,
RJ, 21941-972, Brazil}
\author{J.A. Helayël Neto}
\affiliation{Centro Brasileiro de Pesquisas Físicas (CBPF), 22290-180, Rio de Janeiro, Brazil.}

\begin{abstract}
In this work, we analyze a Lagrangian formalism recently proposed to approach the issue of the Abraham-Lorentz force.
Instead of involving only position and velocity, as usual in Classical Mechanics, this Lagrangian involves the
acceleration of the charge. We find the conserved momentum of the charge in the absence of any field and show
that it contains an acceleration term. This enables us to re-visit the well-known
pre-acceleration problem and show that, contrary to what has been widely believed, it is not related to any violation
of causality.
 \end{abstract}

\maketitle
%
%
%
%

\section{Introduction}

It is well-known that the emission of radiation by an accelerated charge requires a dissipative force to act
in the charged particle to ensure the conservation of energy. Such a force has its origin in the field
produced by the charge, justifying its denomination of a radiation reaction force since, ultimately, it is
a back reaction interaction. For a point charge this dissipative force is given by
the Abraham-Lorentz term\cite{Lorentz}-\cite{Pauli} which is proportional to the third derivative of 
the position of the charge. For a long time this term has provoked a lot of debate in the literature.  
The differential equation of motion including Abraham-Lorentz term 
possess non-physical solutions, such as the runaway solutions for which
the acceleration diverges exponentially\cite{abraham}. We can eliminate these spurious solutions, however
this leads to the charge accelerating before it has suffered any external force, the so-called pre-acceleration\cite{Dirac1938}-\cite{barutpre}
which was met with considerable distrust from scientific community due to its violation of causality.
Solutions to these problems were looked for in many ways. Some researchers have argued that treating the electron
as a point charge is the origin of all the trouble, and that for extended charged bodies we can have solutions
without neither runaway solutions nor causality problems\cite{ford}-\cite{Medina}.

For years, a consistent Lagrangian formulation of 
this problem was tried by researchers, but without success. The power
of such a formulation rests, among other things, in the possibility of readily recognizing who are the 
conjugated momenta and correctly identifying them might shed light on the above mentioned problems. As pointed out by Rohrlich\cite{rohrlich} it is difficult to conciliate the radiation reaction force with the
principle of inertia. He stated that the Abraham-Lorentz equation was incorrect and looked for another equation
of motion free of any inconsistency. O'Connell criticizes\cite{oconnell} Rohlich paper, stating that Abraham-Lorentz
equation is the correct one for a point charge and that Rohrlich equation of motion was exactly the force obtained
for a extended charged body some years before.

 Since the Abraham-Lorentz force involves third derivative, we must work with 
a Lagrangian involving the acceleration of the charge. A recent method available in literature\cite{Galley} is
particularly suitable to obtain the Lagrangian describing a dissipative problem. Recently, it has been
used\cite{galley2} to provide for the first time a consistent Lagrangian formulation to the Abraham-Lorentz force. 
In this paper, we analyze in detail this Lagrangian, showing that it removes any causality problem from
the Abraham-Lorentz force. To do so, we evaluate in this framework the generalized \emph{momenta} for
a charge suffering back-reaction, and we show that this entail a modified momentum conservation equation.  

In order that this report be as self-contained as possible, we dedicate Section II to obtain the Abraham-Lorentz
Lagrangian along the lines developed in Ref.\cite{galley2}. In Section III we evaluate the generalized \emph{momenta},
which enables us in the last Section to re-visit causality issues. 

\section{Lagrangian for the Abraham-Lorentz force}

Let us consider a non-relativistic point charge in the presence of an electromagnetic
field. The Abraham-Lorentz force is given by \cite{Landau}
\begin{equation}
	\mathbf{F}_{RR}=\frac{2e^2\dddot{\mathbf{r}}}{3c^3} \, . \label{frr}
\end{equation}
For limited motions it is easy to verify that the power dissipated by this force compensate for the energy radiated by the charged particle, given
by Larmor formula. The complete equation of motion for the charge is then given by
\begin{equation}
	m\ddot{\mathbf{r}}=e\mathbf{E}+\frac{e}{c}\mathbf{v}\times\mathbf{B}+\mathbf{F}_{RR} \, . \label{lorentzforce}
\end{equation}
Our aim is to obtain the above equation from a consistent Lagrangian formulation. It is clear from (\ref{frr}) that
our Lagrangian must involve the acceleration of the charge. To this end we employ Galley's recent method\cite{Galley},
which consists in doubling the variables of the problem, $\mathbf{r}(t)\rightarrow \{\mathbf{r}_1, \mathbf{r}_2\}$,
imposing $\delta S=0$ to obtain the equations of motion for $\mathbf{r}_{1,2}$ and  finally imposing the 
physical limit $\mathbf{r}_1=\mathbf{r}_2\equiv\mathbf{r}$. 
The action $S$, now a functional of both $\mathbf{r}_1(t)$ and $\mathbf{r}_2(t)$, is given by
\begin{equation}
	S(\mathbf{r}_1,\mathbf{r}_2; t_i;t_f)=\int\limits_{t_i}^{t_f} \Lambda (\mathbf{r}_a, \dot{\mathbf{r}}_a, \ddot{\mathbf{r}}_a)  dt \, , \label{acaogalley}
\end{equation}
with $a=1,2$ and
\begin{eqnarray}
	\Lambda &=&L(\mathbf{r}_1, \dot{\mathbf{r}}_1, \ddot{\mathbf{r}}_1)-L(\mathbf{r}_2, \dot{\mathbf{r}}_2, \ddot{\mathbf{r}}_2)+K(\mathbf{r}_a, \dot{\mathbf{r}}_a, \ddot{\mathbf{r}}_a)\, , \label{Lambda}
\end{eqnarray}
where $K$ is function of both coordinates and cannot be separated. It is particularly relevant for dissipative problems,
such as the one we are dealing with. For more details we refer to the original paper\cite{Galley}. It is useful
to work with the variables $\mathbf{r}_+=(\mathbf{r}_1+\mathbf{r}_2)/2$ and $\mathbf{r}_-=\mathbf{r}_1-\mathbf{r}_2$,
whose physical limit are $\mathbf{r}$ and $\mathbf{0}$, respectively. Imposing $\delta S=0$ we obtain 
the equation of motion from the action given in Eq.(\ref{acaogalley}),
\begin{eqnarray}
	\left[\frac{\partial \Lambda}{\partial \mathbf{r}_-}\right]_{P.L}-\frac{d}{dt}\left[\frac{\partial \Lambda}{\partial \dot{\mathbf{r}}_-}\right]_{P.L}+\frac{d^2}{dt^2}\left[\frac{\partial \Lambda}{\partial \ddot{\mathbf{r}}_-}\right]_{P.L}=0 \, , \label{eqmotion}
\end{eqnarray}
where P.L. denotes the physical limit. The equation for $\mathbf{r}_+$ vanishes in this limit. Employing 
(\ref{Lambda}) it is easy to show that
\begin{eqnarray}
	\left[\frac{\partial \Lambda}{\partial \mathbf{r}_-}\right]_{P.L}=\frac{\partial L}{\partial \mathbf{r}}+\left[\frac{\partial K}{\partial \mathbf{r}_-}\right]_{P.L} \, ,  \label{relentrelek}
\end{eqnarray}
with analogous expressions applying to the terms $\left[\frac{\partial \Lambda}{\partial \mathbf{\dot{r}}_-}\right]_{P.L}$ 
and $\left[\frac{\partial \Lambda}{\partial \mathbf{\ddot{r}}_-}\right]_{P.L}$. Let us now apply this formalism in order
to obtain a lagragian formulation that yields the correct radiation reaction term in the equation of motion. 
In the absence of radiation reaction we have a conservative problem and we have $K=0$. The Lagrangian describing
this situation is well-known in the literature and given by 
\begin{equation}
	L=\frac{1}{2}mv^2-e\phi+\frac{e}{c}\mathbf{v}\cdot\mathbf{A} \, , \label{L}
\end{equation}
where $\phi,\mathbf{A}$ are the electromagnetic potentials describing the external fields. To fully determine
$\Lambda$ in Eq. (\ref{Lambda}) we must determine $K$. After substituting (\ref{L}) into Eq.(\ref{Lambda}) and
the result into Eq.(\ref{eqmotion}), and imposing that the equation of motion must be given by Eq. (\ref{lorentzforce})
we see that $K$ must satisfy 
\begin{equation}
\left[\frac{\partial K}{\partial \mathbf{r}_-}\right]_{P.L}-\frac{d}{dt}\left[\frac{\partial K}{\partial \dot{\mathbf{r}}_-}\right]_{P.L}+\frac{d^2}{dt^2}\left[\frac{\partial K}{\partial \ddot{\mathbf{r}}_-}\right]_{P.L}=\mathbf{F}_{RR} \, .
\end{equation}

One possibility
is to choose 
\begin{equation}
	K=-\frac{2e^2\dot{\mathbf{x}}_-\cdot\ddot{\mathbf{x}}_+}{3c^3} \, . \label{K}
\end{equation}
This completes the construction of the Lagrangian, in accordance with Ref.\cite{galley2}, and a substitution of Eqs. (\ref{L}),(\ref{K}) into
Eq. (\ref{eqmotion}) generates the correct equation of motion for the point charge in the presence
of the external field and suffering a dissipative force given by Abraham-Lorentz term. Observe that there are
other choices of $K$ which would lead us to the same equations of motion. For example, we might also have taken $K'=\frac{2e^2\dot{\mathbf{x}}_+\cdot\ddot{\mathbf{x}}_-}{3c^3} $. One can readily demonstrate that $K$ and $K'$
differes only by a total time derivative. 

\section{The conjugate momentum} 

As it is well-known for Lagrangians containing 
higher derivatives\cite{Barut}, there are two distinct momenta associated to the particle. They are given by
\begin{eqnarray}
	\mathbf{P}^{(1)}_{\pm}&=&\frac{\partial \Lambda}{\partial \dot{\mathbf{r}}_{\mp}}-\frac{d}{dt}\frac{\partial \Lambda}{\partial\ddot{\mathbf{r}}_{\mp}} \label{p1} \\
	\mathbf{P}_{\pm}^{(2)}&=& \frac{\partial \Lambda}{\partial \ddot{\mathbf{r}}_{\mp}} \, . \label{p2}
\end{eqnarray}
where the above definitions ensure that 
in the physical limit only $\mathbf{P}^{(1)}_+$ and $\mathbf{P}^{(2)}_+$ survive.
%
%
Substituting Eqs.(\ref{Lambda}), (\ref{L}) and (\ref{K}) into Eqs.(\ref{p1})-(\ref{p2}) and taking the physical limit
we obtain
\begin{eqnarray}
	\mathbf{P}^{(1)}&=&m\mathbf{v}-\frac{2e^2\mathbf{a}}{3c^3}+\frac{e}{c}\mathbf{A}  \label{P1final}\cr\cr
	\mathbf{P}^{(2)}&=& 0 \, ,
\end{eqnarray}

where $\mathbf{P}^{(1)}$ and $\mathbf{P}^{(2)}$ denote the physical limit of $\mathbf{P}^{(1)}_+$ and $\mathbf{P}^{(2)}_+$
respectively.
As usual, the conjugate momentum of a charge in the presence of the external field is the sum of a mechanical
term and a term containing field variables. The mechanical term, however, is given by
\begin{equation}
	\mathbf{p}^{(1)}=m\mathbf{v}-\frac{2e^2\mathbf{a}}{3c^3} \, , \label{pmec}
\end{equation}
and is now dependent also on the acceleration of the charge, which is not surprising since the Lagrangian
now contains the acceleration.

In terms of these momenta we may find the hamiltonian through the usual Legendre transformation
\begin{eqnarray}
	H&=&\mathbf{P}_+^{(1)}\cdot\dot{\mathbf{r}}_-+\mathbf{P}_-^{(1)}\cdot\dot{\mathbf{r}}_++\mathbf{P}_+^{(2)}\cdot\ddot{\mathbf{r}}_-+\mathbf{P}_-^{(2)}\cdot\ddot{\mathbf{r}}_+-\Lambda \nonumber\\\label{H}
\end{eqnarray}
From the hamiltonian
(\ref{H}), one may readily show\cite{galley2} that
\begin{eqnarray}
	\frac{dE}{dt}=-\frac{\partial L}{\partial t}-\mathbf{\dot{r}}\cdot\left[\frac{d}{dt}\frac{\partial K}{\dot{\mathbf{r}}_-}-\frac{\partial K}{\partial \dot{\mathbf{r}}_-}-\frac{d^2}{dt^2}\frac{\partial K}{\partial \ddot{\mathbf{r}}_-}\right]_{P.L.} \, .
\end{eqnarray}
Observe that in opposition to standard Analytical Mechanics, we can have dissipation of energy even for Lagrangians not
explicitly time dependent. That is the physical role of doubling the variables.
Substituting Eq.(\ref{K}) in the above expression we obtain
\begin{eqnarray}
	\frac{dE}{dt}=\frac{2}{3}\frac{e^2\mathbf{v}\cdot\dddot{\mathbf{r}}}{c^3} \, ,
\end{eqnarray}
which is exactly the power dissipated by the particle, as can be seen by Eq. (\ref{frr}). It can be written
in a more familiar way by rewriting the above expression as
\begin{eqnarray}
	\frac{dE}{dt}=\frac{d}{dt}\frac{2}{3}\frac{e^2\mathbf{v}\cdot\ddot{\mathbf{r}}}{c^3} -\frac{2e^2a^2}{3c^3}\, .
\end{eqnarray}
The first term in the mean does not contribute (considering limited motion) and therefore the power dissipated
is minus the Larmor power, as expected by energy conservation. We turn now to the physical interpretation entailed
by the Lagrangian formalism depicted in this paper.

\section{Causality issues for the Abraham-Lorentz force} 

Translational symmetry of a system entails momentum conservation, as it is known 
since Noether's theorem\cite{Goldstein}. For Lagrangians involving only positions and velocities of
the particles the conserved momentum is given by
\begin{equation}
\mathbf{p}=\sum_{\alpha}\frac{\partial L}{\partial \mathbf{\dot{r}}} \, ,
\end{equation}
where $\alpha$ indexes the particles. Naturally, when the Lagrangian involves higher-order derivatives
the expression for the conserved momentum must change. Let us now find its expression. Suppose that
we have a Lagrangian $\Lambda(\mathbf{r}_+,\mathbf{r}_-,\mathbf{\dot{r}}_{+}, \mathbf{\dot{r}}_{-},\mathbf{\ddot{r}}_{+},\mathbf{\ddot{r}}_{-})$. Translational symmetry means that the the Lagrangian do not change if we perform the
changes $\mathbf{r}_+\rightarrow \mathbf{r}_++\delta\mathbf{r}$ and $\mathbf{r}_-\rightarrow \mathbf{r}_-+\delta\mathbf{r}$.
In such a way we must have
\begin{equation}
\left(\frac{\partial \Lambda}{\partial \mathbf{r}_+}+\frac{\partial \Lambda}{\partial \mathbf{r}_-}\right)\delta\mathbf{r}=0
\end{equation}
Employing the equations of motion we recast the last equation in the form
\begin{equation}
\left[\frac{d}{dt}\left(\frac{\partial \Lambda}{\partial \mathbf{\dot{r}}_+}-\frac{d}{dt}\frac{\partial \Lambda}{\partial \mathbf{\ddot{r}}_+}+\frac{\partial \Lambda}{\partial \mathbf{\dot{r}}_-}-\frac{d}{dt}\frac{\partial \Lambda}{\partial \mathbf{\ddot{r}}_-}\right)\right]\delta\mathbf{r}=0 \, .
\end{equation}
In terms of the definition (\ref{p1}) we see that the momentum $\mathbf{P}_+^{(1)}+\mathbf{P}_-^{(1)}$ must be conserved.
In the physical limit only $\mathbf{P}_+^{(1)}$ survives and is given in Eq.(\ref{P1final}). 
In the absence of external fields the problem of a point charge has translational symmetry and thus
we see that its mechanical momentum $\mathbf{p}^{(1)}$ given in Eq.(\ref{pmec}) must be conserved.
Note that the equation of motion (\ref{eqmotion}) can be recast in the form
\begin{equation}
			\frac{d\mathbf{p}^{(1)}}{dt}=e\mathbf{E}+\frac{e}{c}\mathbf{v}\times\mathbf{B}\, . \label{dp1dt}
\end{equation}

Naturally, we could have written Eq. (\ref{dp1dt}) directly from (\ref{lorentzforce}). However, what the 
Lagrangian formalism allows us to realize is that the LHS of Eq.(\ref{dp1dt}) is indeed the variation rate of the momentum
of the particle. The law of inertia may be conceived as the conservation of moment for an isolated particle.
Our results show that, in the absence of external fields, what is conserved is not the velocity, but the moment 
$\mathbf{p}^{(1)}$. In spite of the initial discomfort such a claim may provoke, we would like to remember that
all Newtonian Mechanics is based on Lagrangians that depend only on coordinates and velocities. This is related
to the fact that in Newtonian Mechanics all interactions are supposed to be instantaneous. In those cases, the
trajectory of the particles are determined only by their initial positions and velocities, which evidently is
not the case when we have radiation reaction. Therefore, it is not surprising that some basic tenets of Newtonian
Mechanics must be thought anew when we allow for interactions propagating with finite velocity. The Lagrangian
formalism developed in this work shows that, if we are to think of force as the variation rate of momentum (now given
by Eq.(\ref{pmec})), we must modify the law of inertia. In Ref.\cite{rohrlich}, Rohrlich argued that the law of
inertia in its usual form and the Abraham-Lorentz force can not be conciliated. In his paper he tried to
change the Abraham-Lorentz force, while here we show that all the confusion arose from an erroneous law of conservation.
Phenomena involving pre-acceleration, where the charge begins to accelerate before the action of an external field, 
are to be expected when the radiation reaction is accounted for, as showed by Dirac\cite{Dirac1938}. However,
the Lagrangian formalism presented here enables us to see that
there is no break of causality involved here. Indeed, nothing prevents a charge from being accelerated in the absence
of fields. We wish to emphasize that this does not violate the symmetry of space. Contrary, in the Lagrangian
formalism the conservation of momentum is a consequence of this symmetry. Everything depends upon the initial conditions.

To conclude, and we shall be reporting on that elsewhere, we highlight the potentiality of the method presented in
\cite{Galley} to yield an expression for the planar ((1+2) dimensions) of the Larmor expression for the power irradiated
by an accelerated charge. The problem is highly non-trivial, in view of the support of the retarded Green's function
for the planar electromagnetic field, but the method of Ref.\cite{Galley} may show to be extremely helpful.

\noindent
{\bf Acknowledgements}\\
The authors are indebted with Carlos Farina and Thiago Hartz for
valuable discussions. The authors  also  thank to CNPq and FAPERJ (brazilian agencies) for
partial financial support.

\end{document}